\documentclass[preprint,showkeys,secnumarabic,showpacs,amsmath,amssymb]{revtex4}
\usepackage{graphicx}
\begin{document}
\title{Curvaton Reheating and Intermediate Inflation in Brane Cosmology}
\author{Hossein Farajollahi}
\email{hosseinf@guilan.ac.ir} \affiliation{Department of Physics,
University of Guilan, Rasht, Iran}
\author{Arvin Ravanpak}
\email{aravanpak@guilan.ac.ir} \affiliation{Department of Physics,
University of Guilan, Rasht, Iran}

\date{\today}

\begin{abstract}
In this paper, we study the curvaton reheating mechanism for an intermediate inflationary universe in brane world
cosmology. In contrast to our previous work, we assume that when the universe enters the kination era, it is still in the
high-energy regime. We then discuss, in detail, the new cosmological constraints on both the model parameters and the
physical quantities.

\end{abstract}


\maketitle

\section{Introduction}

Inflation as the most promising framework for
understanding the physics of the very early universe, relates the
universe evolution to the properties of one or more scalar
inflaton fields that is responsible for creating universe acceleration.
This in turn causes a flat and homogeneous universe which later evolves into the
present structure. Furthermore,
inflation can be a solution to many other cosmological issues such as horizon
and flatness problems \cite{Tzirakis}--\cite{Guth}. The inflationary universe
model may also provide a causal explanation of the origin of
the observed anisotropy of the cosmic microwave background radiation (CMBR), and also large scale structures distribution
\cite{Albrecht}--\cite{Hinshaw}.

In intermediate inflation, the scale factor
of the universe in a certain extend intermediate between a power law
and exponential expansion whereas the expansion rate
is faster than power law and slower than exponential ones. In this kind, the slow-roll
approximation conditions are satisfied with time and thus same as in power-law inflation case, within the
model, there is no natural end to inflation \cite{Rendall}--\cite{Liddle}. While during the inflationary period the
universe becomes dominated by the inflation scalar potential, at
the end of inflation the universe represents a combination of kinetic
and potential energies related to the scalar field, which is
assumed to take place at very low temperature\cite{Randall}--\cite{Campuzano}.

In the standard reheating mechanism after inflation while the temperature grows in many orders of magnitude, most of the matter and
radiation of the universe was created via the decay of the inflaton
field and the Big-Bang universe is recovered. The reheating temperature is particularly very important to cosmologists. In the standard mechanism of reheating, in the stage of
 oscillation of the scalar field a minimum in the
inflaton potential is something crucial for the reheating mechanism \cite{Kofman}--\cite{del}.

However, since the scalar field potential in some models does not present a minimum, the conventional
mechanism, offered to bring inflation to an end, becomes ineffective.
These models are well known as non-oscillating or simply NO Models.
To overcome this shortcoming, a scalar field, dubbed curvaton is represented in the mechanism of
reheating in these kind of models \cite{Mollerach}--\cite{Lyth}. The curvaton scenario is suggested as an alternative to produce the primordial scalar perturbation where in turn is responsible for the
 structure formation. When it decay into conventional matter, it creates an efficient
mechanism of reheating. The energy density of carvaton field is not diluted during inflation therefore the curvaton may be
responsible for some or all the matter content of the universe at the
present time \cite{Papantonopoulos,Campuzano}.

In the Randall-Sundrum cosmology, the issue of inflation of a single brane in
an AdS bulk successfully consolidates the idea
that universe is in a three-dimensional brane within a
higher-dimensional bulk spacetime. In the high energy limit, the brane
inflationary models contain correction terms
in their Friedmann equations with significant consequences in the inflationary dynamics \cite{Binetruy}--\cite{Shir}. In this work, in the context of brane intermediate inflationary cosmology, we would like to introduce
the curvaton field as a mechanism to bring inflation to
an end. A similar approach to study the subject has been investigated in \cite{del} and \cite{Lopez}. The authors in
\cite{del}, employs curvaton scenario for intermediate inflationary models
in standard cosmology whereas in \cite{Lopez}, the steep inflation model
in brane cosmology has been studied with the curvaton field responsible to bring inflation to an end. Here, in an attempt to integrate the above two works, curvaton reheating and intermediate inflation in brane cosmological models has been investigated with the aim to explore the dynamics of the universe within and after inflation era. In \cite{lowenergy}, we assumed that when the universe enters the kination era, it is in the low energy regime. Here, in both inflationary and kination epochs the universe is in high energy regime. The model then suggests new constraints on the space parameters and predict reheating temperature in different scenarios.

\section{The Model}

In this section, a five-dimensional brane cosmology is studied with the
modified Friedmann equation given by \cite{Shiromizu,Kamenshchik},
\begin{equation}\label{fried}
H^2=\kappa\rho_\phi[1+\frac{\rho_\phi}{2\lambda}]+\frac{\Lambda_4}{3}+\frac{\xi}{a^4},
\end{equation}
where $H = \dot{a}/a$ is the Hubble parameter. The $\rho_\phi$ and $\Lambda_4$ are,respectively, the matter field confined to the
brane and the four-dimensional cosmological constant. we
also assume that $\kappa = 8\pi G/3 = 8\pi/(3m_p^2)$. The influence of the bulk gravitons on the brane is shown in the last term of the equation, where $\xi$ is
integration constant. The four and five-dimensional Planck masses are related through brane tension $\lambda$ in $m_p
= \sqrt{3M_5^6/(4\pi\lambda)}$, where constrained by nucleosynthesis. The brane tension
satisfies the inequality $\lambda
> (1MeV)^4$. In the following, in the inflation epoch, we assume that the universe is in the high energy
regime, i.e. $\rho_\phi\gg\lambda$. We suppose that the four-dimensional cosmological constant is
vanished. In addition, at the beginning of inflation, the last term in (\ref{fried}) vanishes and, thus we are left with the effective Friedmann equation given by
\begin{equation}\label{efffried}
H^2\simeq\beta\rho_\phi^2,
\end{equation}
where $\beta$ is equal to $\kappa/(2\lambda)$, with the dimension of $m_p^{-6}$. Furthermore, we assume that the inflaton field is bounded to the brane, therefore
its field equation is in the form of:
\begin{equation}\label{scalarfield}
\ddot{\phi}+3H\dot{\phi}+V'=0.
\end{equation}
In the above equation, $V(\phi)$ is the effective
scalar potential. The dot and prime respectively represent derivative with respect to the
cosmological time and scalar
field $\phi$. We also have the conservation equation as
\begin{equation}
\dot{\rho_\phi}+3H(\rho_\phi+p_\phi)=0,
\end{equation}
where the energy density and pressure respectively are given by $\rho_\phi=(\dot{\phi}^2/2)+V(\phi)$, and
$p_\phi=(\dot{\phi}^2/2)-V(\phi)$. For convenience, we also take units in which
$c={\hbar}=1$.

In intermediate inflationary scenarios, one can find the exact solution by taking the scale factor $a(t)$ as
\begin{equation}\label{inter}
a(t)=\exp(At^f),
\end{equation}
where the power $f$ is a parameter limited as $0<f<1$, and the coefficient $A$ is a
positive constant with dimension $m_P^f$.

From equations (\ref{efffried}), (\ref{scalarfield}) and (\ref{inter}), one can find potential
$V(\phi)$ and scalar field $\phi(t)$ as
\begin{equation}
V(\phi)={\frac{2(1-f)}{9\beta}}[6fA{(\frac{3\beta^{1/2}}{4(1-f)})^f}\phi^{2(f-1)}-\frac{(1-f)}{\phi^2}],
\end{equation}
and
\begin{equation}\label{phi}
\phi(t)=[{\frac{4(1-f)}{3\beta^{1/2}}}t]^{1/2}.
\end{equation}
Therefore, we obtain the Hubble parameter in terms of inflaton scalar field $\phi$
as
\begin{equation}\label{h}
H(\phi)=fA(\frac{3\beta^{1/2}}{4(1-f)})^{f-1}\phi^{-2(1-f)}.
\end{equation}
By solving the field equations in slow roll approximation, with simple power law scalar potential in the form of
\begin{equation}
V(\phi)=fA\beta^{(f-2)/2}[\frac{3}{4(1-f)}]^{f-1}\phi^{-2(1-f)}\label{pot},
\end{equation}
One can find scale factor sam as in equation (\ref{inter}). With the potential (\ref{pot}) in the slow roll approximation we find solutions for $\phi(t)$ and $H(\phi)$.
these solutions are similar to those obtained in
the exact solution and expressed by (\ref{phi}) and (\ref{h}).
It is notable that no extreme point can be found for these potentials.

In our model, the dimensionless slow roll parameters $\varepsilon\simeq V'^2/(3\beta V^3)$ and $\eta\simeq V''/(3\beta
V^2)$, respectively, become
\begin{equation}
\varepsilon\simeq\frac{4(1-f)^2}{3fA\beta^{f/2}[\frac{3}{4(1-f)}]^{f-1}}\phi^{-2f},
\end{equation}
and
\begin{equation}
\eta\simeq\frac{2(1-f)(3-2f)}{3fA\beta^{f/2}[\frac{3}{4(1-f)}]^{f-1}}\phi^{-2f}.
\end{equation}
Note that for $0<f<1$, the ratio $\varepsilon/\eta=2(1-f)/(3-2f)$ is less than one. Moreover, $\eta$ earlier than $\varepsilon$ reaches unity. As a result, the end of inflation is controlled by the condition $\eta=1$ instead of $\varepsilon=1$. We also find the inflaton field at
the end of inflation as
\begin{equation}\label{phie}
\phi_e=[\frac{2(1-f)(3-2f)}{3fA\beta^{f/2}[\frac{3}{4(1-f)}]^{f-1}}]^{1/2f},
\end{equation}
where the subscript $"e"$ is used to indicate end of inflationary period.

\section{The Curvaton Field Constraint During The Kinetic Epoch}

With disregarding the term $V'$ in equation (\ref{scalarfield}) compare to the friction term
$3H\dot{\phi}$, the model enters a new era called the kinetic epoch or kination. From now on, we use
the subscript (or superscript) $"k"$, to mark quantities
at the beginning of this period. During the kination era
we have $\dot{\phi}^2/2 > V(\phi)$ which could be seen as a stiff
fluid, since $p_\phi=\rho_\phi$. Since we assume that the universe in the kinetic epoch is still in high energy regime, the field equations (\ref{fried}) and (\ref{scalarfield}) become,
$H^2=\beta\dot{\phi}^4/4$ and $\ddot{\phi}+3H\dot{\phi}=0$, where
the second equation gives
\begin{equation}
\dot{\phi}={\dot{\phi}}_k(\frac{a_k}{a})^3.
\end{equation}
Therefore, energy density and Hubble parameter in terms of scale factor respectively become
\begin{equation}\label{rhophi}
\rho_\phi(a)=\rho_\phi^k(\frac{a_k}{a})^6,
\end{equation}
and
\begin{equation}\label{hk}
H(a)=H_k(\frac{a_k}{a})^6.
\end{equation}
In the above equations the "k" label is associated to the inflaton field at the beginning
of the kinetic era.

With the curvaton field complying the Klein-Gordon equation, we assume
that the scalar potential associated to the curvaton is given by
$U(\sigma)=m^2\sigma^2/2$, with $m$ to be curvaton mass.

In comparison with the curvaton energy density, let us consider that $\rho_\phi$ is dominant component. Further, assume that curvaton field oscillates around minimum of its
effective potential. During kination era, the curvaton density
evolves as a non-relativistic matter, i.e. $\rho_\sigma\propto
a^{-3}$ and inflaton stays dominated. Therefore, by decay of the the curvaton
field into radiation, the Big-Bang cosmology is recovered.

In the inflation epoch, the curvaton field is assumed to be effectively massless. Therefore, the curvaton rolls down
its potential until its kinetic energy is weakened by exponential expansion and then almost vanishes.
As a result the curvaon field becomes frozen roughly to a constant value, $\sigma_\ast\approx\sigma_e$. The
subscript $"\ast"$ denotes the period when the cosmological scale
exits the horizon.

We also assume that during kination era, the
Hubble parameter value decreases and become equal to the
curvaton mass, i.e. $m\simeq H$. Then, at this point, curvaton
field becomes effectively massive and from equation (\ref{hk}), we
get
\begin{equation}\label{ratiom}
\frac{m}{H_k}=(\frac{a_k}{a_m})^6,
\end{equation}
where the subscript $"m"$ marks quantities during kination at the time when the
curvaton mass $m$ has order of $H$.

To prohibit a period of curvaton-driven inflation, the inflaton domination must still hold, i.e.
$\rho_\phi^m\gg\rho_\sigma$($\sim U(\sigma_e)\simeq
U(\sigma_\ast))$. This inequality gives us a constraint on
the values of the curvaton field $\sigma_\ast$. When
$H\simeq m$,then, we get $\frac{m^2\sigma_\ast^2}{2\rho_\phi^m}\ll1$,
meaning that the curvaton field $\sigma_\ast$ satisfies the
constraint
\begin{equation}\label{sigmaastsquare}
\sigma_\ast^2\ll\frac{2}{m\beta^{1/2}} \cdot
\end{equation}
the ratio of the potential energies at the end of inflation,
becomes
\begin{equation}
\frac{U_e}{V_e}=\frac{m^2\sigma_\ast^2\beta^{1/2}}{2H_e}<1,
\end{equation}
and, therefore, curvaton energy sub dominated. The curvaton mass then should obey the constraint
\begin{equation}\label{mh}
m^2<H_e^2=(fA)^{2/f}(\frac{3-2f}{2})^{2(f-1)/f}.
\end{equation}
The above constraint imposed by the fact that the curvaton field during the inflationary era must be
effectively massless and therefore we have $m<H_e$. When curvaton field mass becomes important,
i.e. $m\simeq H$, its energy density decays same as a
non-relativistic matter as
$\rho_\sigma=m^2\sigma_\ast^2a_m^3/(2a^3)$. In fact whenever
curvaton undergoes quasi-harmonic oscillations, the
potential and kinetic energy densities become comparable.

In general, the curvaton field decay happens in two different scenarios. When the curvaton field first dominates the cosmic
expansion and then decays. Or its decay occurs before its domination. Next section we discusses these scenarios in more details.

\section{Constraints on model parameters}

{\bf Case 1: Curvaton decay after domination}

If the curvaton field dominates the cosmic expansion (i.e.
$\rho_\sigma>\rho_\phi$), then, at a distance, let us assume that, when $a=a_{eq}$, the
inflaton and curvaton energy densities become equal. Therefore, from equations (\ref{rhophi}) and (\ref{hk}), by using $\rho_\sigma\propto a^{-3}$, we obtain
\begin{equation}\label{ratiophi}
(\frac{\rho_\sigma}{\rho_\phi})|_{a=a_{eq}}=\frac{\beta^{1/2}m^2\sigma_\ast^2}{2}
\frac{a_m^3}{a_k^6}\frac{a_{eq}^3}{H_k}=1.
\end{equation}
Using (\ref{hk}), (\ref{ratiom}) and (\ref{ratiophi}), the Hubble
parameter, $H_{eq}$, in terms of the curvaton parameters can be obtained
\begin{equation}\label{hableeq}
H_{eq}=H_k(\frac{a_k}{a_{eq}})^6=\beta m^3\sigma_\ast^4/4.
\end{equation}
The decay parameter $\Gamma_\sigma$ is bounded by
nucleosynthesis, so the curvaton field must decay
before nucleosynthesis which means that
$H_{nucl}\sim10^{-40}<\Gamma_\sigma$ (in units of Planck mass
$m_p$). From $\Gamma_\sigma<H_{eq}$, we obtain a constraint on $\Gamma_\sigma$, as
\begin{equation}\label{10-40}
10^{-40}<\Gamma_\sigma<\beta m^3\sigma_\ast^4/4.
\end{equation}
Following the argument given in \cite{del}, by studying the scalar perturbations related to the
curvaton field $\sigma$, we find new constraints on our model parameters. When the curvaton field decays, the Bardeen parameter, $P_\zeta$,
whose observed value is about $2.4\times 10^{-9}$ \cite{Komatsu}, becomes \cite{Lyth}
\begin{equation}\label{bardeen0}
P_\zeta\simeq\frac{1}{9\pi^2}\frac{H_\ast^2}{\sigma_\ast^2}\cdot
\end{equation}
With regards to $\sigma_\ast^2\gg H_\ast^2/(4\pi^2)$, the spectrum of fluctuations is automatically gaussian and $\Gamma_\sigma$ independent, thus, one can ease the analysis of space
parameter. From expressions (\ref{h}), (\ref{phie}) and (\ref{bardeen0}), we may write
\begin{equation}\label{Aafter}
A=[\frac{9\pi^2\sigma_\ast^2P_\zeta}{f^2}(\frac{3-2f}{2f}-N_\ast)^{\frac{2(1-f)}{f}}]^{f/2},
\end{equation}
where
\begin{equation}\label{Nstar}
N_\ast=\int_{t_\ast}^{t_e}{H(t')dt'}=A[\frac{3\beta^{1/2}}{4(1-f)}]^f(\phi_e^{2f}-\phi_\ast^{2f}).
\end{equation}
Equation (\ref{Nstar}) defines number of e-folds corresponding to the cosmological
scales, i.e. the number of remaining inflationary e-folds at the
time when the cosmological scale exits the horizon. Then, the constraint given by equation (\ref{mh}) changes to
\begin{equation}\label{m2}
m^2<9\pi^2\sigma_\ast^2P_\zeta(1-\frac{2fN_\ast}{3-2f})^{2(1-f)/f}.
\end{equation}
By using equations (\ref{10-40}) and (\ref{m2}) we obtain an upper limit on $\Gamma_\sigma$ as
\begin{equation}\label{gammaafter}
\Gamma_\sigma<\frac{27\pi^3}{4}\beta\sigma_\ast^7P_\zeta^{3/2}(1-\frac{2fN_\ast}{3-2f})^{3(1-f)/f}.
\end{equation}
Moreover, by using
Big Bang Nucleosynthesis $(BBN)$ temperature, $T_{BBN}$, we give the constraints for our model parameters $A$ and $f$.
The reheating temperature is given by $T_{BBN}\sim10^{-22}$ (in unit of $m_p$). Since the reheating temperature occurs before the $BBN$, $T_{reh}>T_{BBN}$, by knowing that $T_{reh}\sim\Gamma_\sigma^{1/2}>T_{BBN}$ one obtains the following constraints,
\begin{equation}\label{hast}
H_\ast^2=f^2A^{2/f}[\frac{3-2f}{2f}-N_\ast]^{2(f-1)/f}>(36\pi^2)^{4/7}
(P_\zeta T_{BBN})^{4/7}\sim10^{-16},
\end{equation}
where for later consideration we have used the scalar spectral index $n_s$ closed to one and thus $m\approx H_*/10$ \cite{Sanchez}.

By fitting equation (\ref{hast}) in its lower
limit, Fig. 1 (left panel) shows contours of model parameters $f$ and $A$ corresponding to different number of e-folds, $N_\ast$. Similarly, Fig. 1 (right panel) shows the constraints when the upper limit, $H_\ast\leqslant10^{-5}$ \cite{Dimopoulos}, is used.

\begin{figure}[h]
\centering
\includegraphics[scale=0.35]{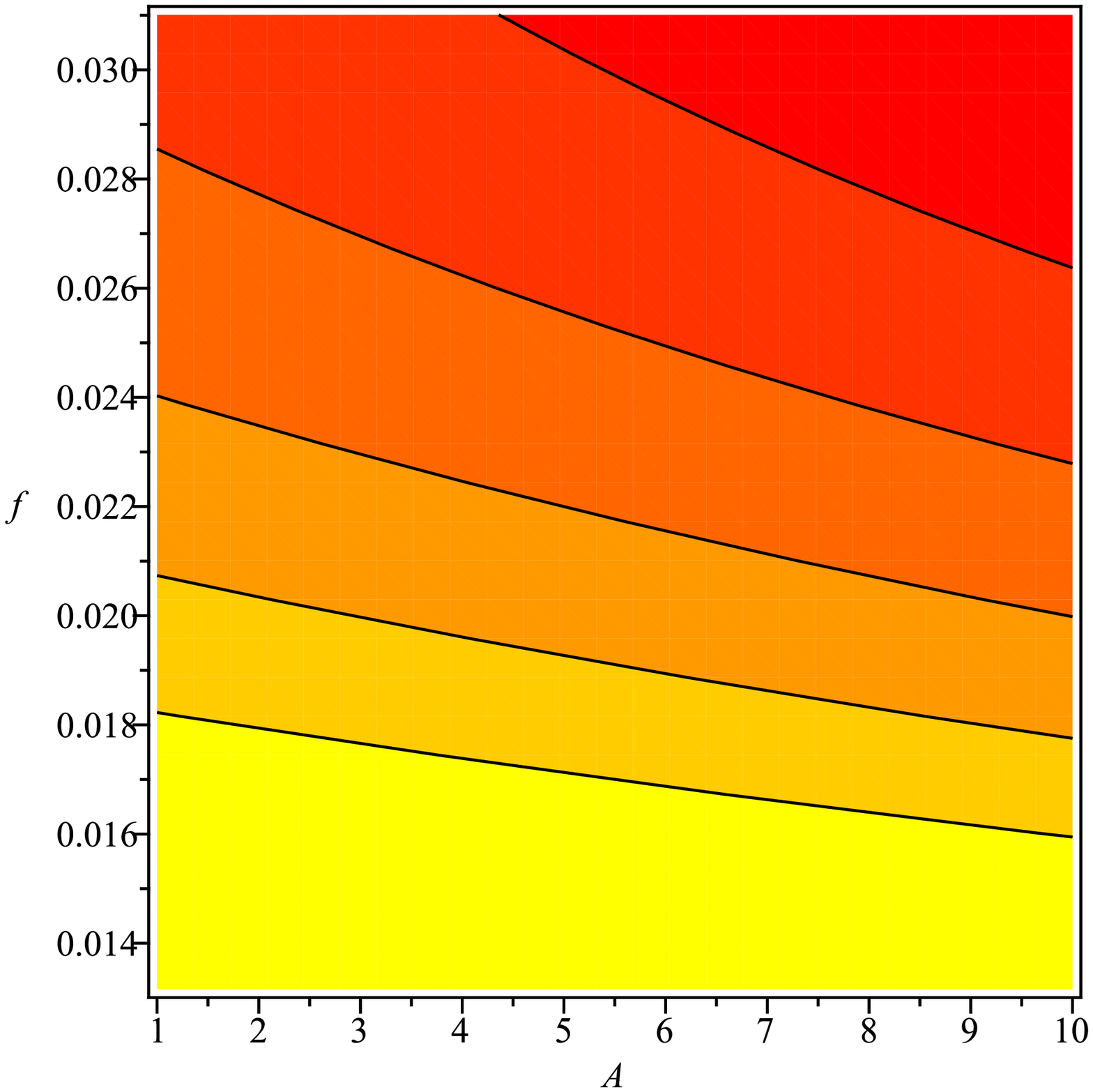}\includegraphics[scale=0.35]{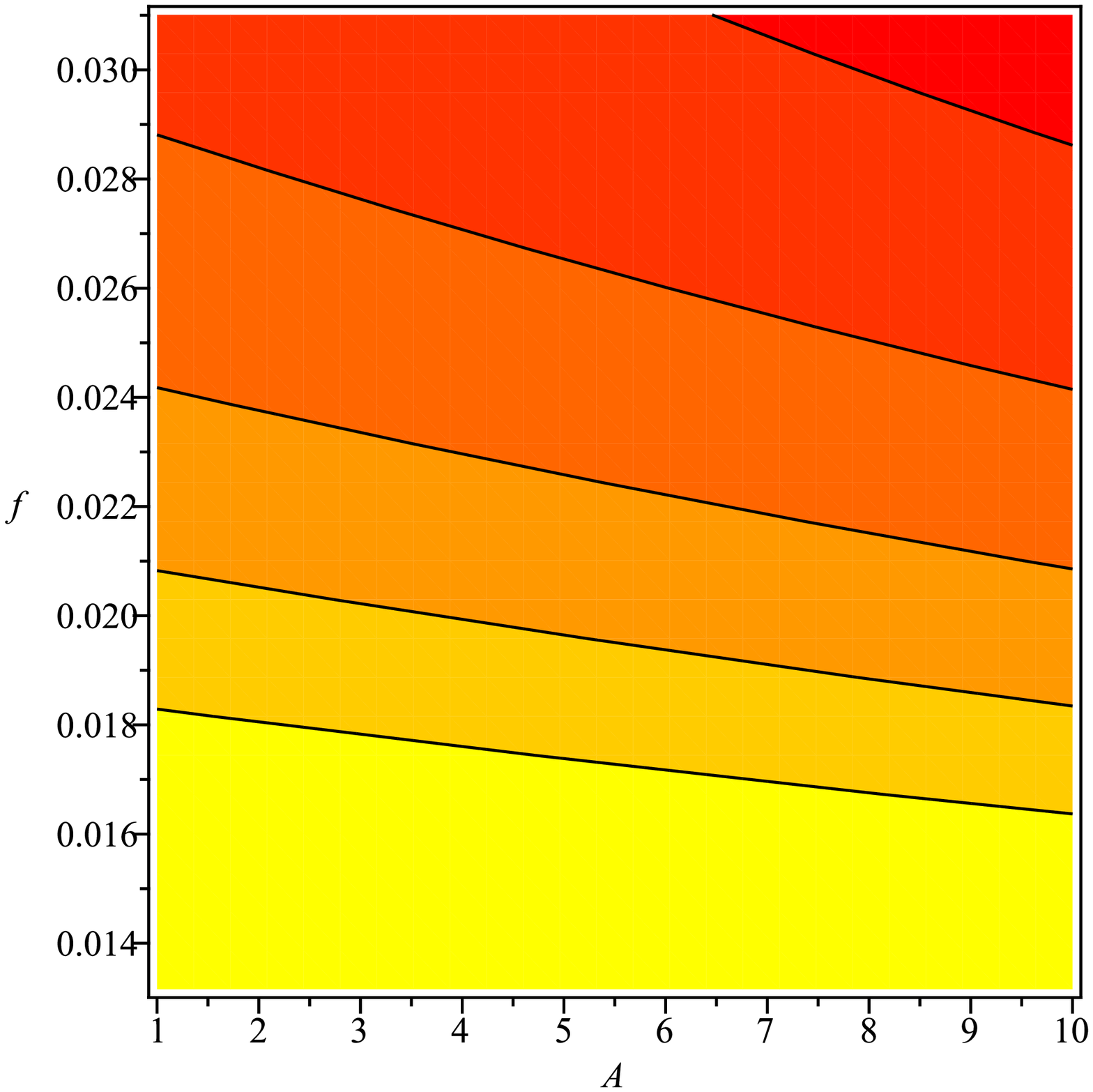}
\caption{contour plot for the parameters
$f$ and $A$ in both curvaton decay after and before domination, fitted from the lower limit of the $T_{BBN}$ (left panel) and upper limit of the
$T_{BBN}$ (right panel). Lower values of the $N_\ast$
parameters correspond to darker regions and the contour levels are separated by the quantity
$\bigtriangleup N_\ast=10$. Here, we have considered the values of $N_\ast$, from $40$ to $80$.}
\end{figure}

{\bf Case 2: Curvaton decay before domination}

We assume that the curvaton decays before its energy
density becomes greater than the inflaton energy density. Furthermore, the
mass of the curvaton is considered toe comparable with the Hubble
expansion rate, such that $\rho_\sigma\propto
a^{-3}$. Following \cite{del}, we also assume that the curvaton field decays when
$\Gamma_\sigma= H(a_d)=H_d$. Then, from equation (\ref{hk}) we obtain
\begin{equation}\label{gammasig}
\Gamma_\sigma=H_d=H_k(\frac{a_k}{a_d})^6,
\end{equation}
where "d" stands for quantities when the curvaton decays.

By assuming that $\Gamma_\sigma<m$ and $\Gamma_\sigma>H_{eq}$), we yield,
\begin{equation}\label{kappa}
\frac{\beta m^2\sigma_\ast^4}{4}<\frac{\Gamma_\sigma}{m}<1.
\end{equation}
In this scenario, the Bardeen parameter becomes
\begin{equation}\label{bardeen}
P_\zeta\simeq\frac{r_d^2}{16\pi^2}\frac{H_\ast^2}{\sigma_\ast^2},
\end{equation}
where $r_d=\frac{\rho_\sigma}{\rho_\phi}|_{a=a_d}$. Employing
$\rho_\sigma=m^2\sigma_\ast^2a_m^3/(2a^3)$ and simple algebraic manipulation in (\ref{rhophi}), (\ref{ratiom}) and (\ref{gammasig}) we get
\begin{equation}\label{r-d}
r_d=\frac{\sigma_\ast^2}{2}(\frac{\beta m^3}{\Gamma_\sigma})^{1/2}.
\end{equation}
From expressions (\ref{h}), (\ref{phie}), (\ref{Nstar}), (\ref{bardeen}) and (\ref{r-d}), one yields
\begin{equation}\label{Abefore}
A=[\frac{64\pi^2P_\zeta\Gamma_\sigma}{\beta m^3\sigma_\ast^2f^2}(\frac{3-2f}{2f}-N_\ast)^{\frac{2(1-f)}{f}}]^{f/2}.
\end{equation}
In addition, the constraint given by equation (\ref{mh}) becomes
\begin{equation}\label{m4}
m^2<\frac{64\pi^2P_\zeta\Gamma_\sigma}{\beta m^3\sigma_\ast^2}(1-\frac{2fN_\ast}{3-2f})^{2(1-f)/f}.
\end{equation}
Now, by using equations (\ref{kappa}) and (\ref{m4}), we could get the inequality
\begin{equation}\label{gammabefore}
\Gamma_\sigma<\frac{64\pi^2
P_\zeta}{\beta m^3\sigma_\ast^2}(1-\frac{2fN_\ast}{3-2f})^{2(1-f)/f},
\end{equation}
which gives an upper limit on $\Gamma_\sigma$.

Finally, since the reheating temperature satisfies
$T_{reh}>T_{BBN}$, and also $\Gamma_\sigma>T_{BBN}^2$, we obtain,
\begin{equation}\label{hast2}
H_\ast^2=f^2A^{2/f}[\frac{3-2f}{2f}-N_\ast]^{2(f-1)/f}>(64\pi^2)^{4/7}(P_\zeta
T_{BBN})^{4/7}\sim10^{-16},
\end{equation}
where, as before, we have used the scalar spectral index $n_s$
closed to one. Note some similarities between (\ref{hast2}) and (\ref{hast}), and therefore the same constraints on $A$ and $f$ for different number of e-folds can be considered in this case.

\section{Constraints on Curvaton Mass}

By using tensor perturbations scheme, one can constrain the curvaton mass. From \cite{Kamenshchik}, the gravitational wave amplitude is given by,
\begin{equation}\label{hgw}
h_{GW}\simeq C_1H_\ast,
\end{equation}
where $C_1$ is a constant. If inflation occurs at an energy scale smaller than the grand
unification, we have $H\ll10^{-5}$, an advantage of curvaton approach over single inflaton field approach. Then, using $H_\ast^2=\beta V_\ast^2$, and (\ref{hast2}) we find
\begin{equation}\label{grav-wave}
h_{GW}^2\simeq C_1^2f^2A^{2/f}[\frac{3-2f}{2f}-N_\ast]^{2(f-1)/f}.
\end{equation}
From equations (\ref{mh}) and (\ref{grav-wave}), we yield the constraint
\begin{equation}
m^2<\frac{h_{GW}^2}{C_1^2}(\frac{3-2f-2fN_\ast}{3-2f})^{2(1-f)/f},
\end{equation}
which gives an upper limit for the curvaton mass.

The authors in \cite{Langlois} constrain the density fraction of the gravitational
wave as
\begin{equation}\label{I}
I\equiv
h^2\int_{k_{BBN}}^{k_\ast}\Omega_{GW}(k)dlnk\simeq2h^2\epsilon\Omega_\gamma(k_0)h^2_{GW}
(\frac{H_\ast}{\widetilde{H}})^{2/3}\leq2\times 10^{-6},
\end{equation}
where $k_{BBN}$ is physical
momentum corresponding to the horizon at $BBN$. Also, $\Omega_{GW}(k)$ is density fraction of the gravitational
wave as a function of $k$. At the present epoch we have $\Omega_\gamma(k_0)=2.6\times10^{-5}h^{-2}$, $\epsilon\sim10^{-2}$ and $h=0.73$. Therefore, for the curvaton decay after or before domination, $\widetilde{H}$
is either $\widetilde{H}=H_{eq}$ or $\widetilde{H}=H_d$.

{\bf Case 1: Curvaton decay after domination}

By using  equations (\ref{hableeq}), (\ref{bardeen0}) and (\ref{hgw}),
the constraint on (\ref{I}) becomes
\begin{equation}\label{m-sigma}
m^3\gtrsim7.5\times10^{-37}{P_\zeta}^2\sim10^{-54}.
\end{equation}
If the decay rate is of gravitational strength, then $\Gamma_\sigma\sim m^3$ \cite{Rodriguez}--\cite{Lazarides}
and equation (\ref{10-40}) imposes a third constraint as
\begin{equation}\label{msquare1}
\sigma_\ast^4>\frac{4}{\beta}\simeq7.55\times10^{-25}.
\end{equation}
We also have two constraints (\ref{sigmaastsquare}) and (\ref{10-40}). By using
equations (\ref{mh}) and (\ref{bardeen0}) for the curvaton decay after domination we find
\begin{equation}\label{ec}
\frac{m^2}{\sigma_\ast^2}<9\pi^2P_\zeta\sim 2.13\times10^{-7}.
\end{equation}
Therefore, for the
curvaton decay after domination, we have five constraints (\ref{sigmaastsquare}), (\ref{10-40}), (\ref{m-sigma})--(\ref{ec}) on $m$ or $\sigma_\ast$. Fig. 2 (left panel), shows the allowed region for curvaton mass in shaded color.

{\bf Case 2: Curvaton decay before domination}

In this case, using the constraint on the density fraction of the gravitational
wave, equation (\ref{I}), and also using equations (\ref{gammasig}), (\ref{bardeen}) and (\ref{r-d}), with regards to $\Gamma_\sigma^{1/2}>T_{BBN}$ we obtain
\begin{equation}\label{m-sigmastar}
m^3\sigma_\ast^2\gtrsim
1.3\times10^{-30}P_\zeta T_{BBN}\sim10^{-61}.
\end{equation}
By incorporating equations (\ref{kappa}) and (\ref{m-sigmastar}), a second constraint is given by \cite{Herrera}
\begin{equation}\label{msquare}
m^2>1.5\times10^{-18}P_\zeta T_{BBN}\sim10^{-49}.
\end{equation}
By employing equation (\ref{kappa}), a pair of new constraints can also be obtained as
\begin{equation}\label{msquare2}
m^2<1,
\end{equation}
and
\begin{equation}\label{msquare3}
\sigma_\ast^4<7.55\times10^{-25}.
\end{equation}
Moreover, from equation (\ref{m-sigmastar}) one can get a further constraint
\begin{equation}\label{sigmam5/4}
m^{3/2}\sigma_\ast^2>\frac{7\times10^{-6}P_\zeta}{\beta}\simeq3.17\times10^{-39}.
\end{equation}
For curvaton decay before domination, we have the constraint (\ref{sigmaastsquare}).
Further, from (\ref{mh}), (\ref{kappa}), (\ref{bardeen}) and (\ref{r-d}), a second constraint is 
\begin{equation}\label{ec2}
m^4\sigma_\ast^2<1.2\times10^{-23}\pi^2P_\zeta\sim 2.9\times10^{-31}.
\end{equation}
Therefore, there are seven constraints (\ref{sigmaastsquare}), (\ref{m-sigmastar})--(\ref{ec2}) on either $m$ or $\sigma_\ast$, for the
curvaton decay before domination. In Fig. 2 (right panel), the shaded region shows the allowed region for $m$ which is bounded by the above constraints. Here, we have fixed the value of $\lambda$
(the tension of the brane) by $\lambda=7.9\times10^{-25}$ (in units of $m_p$)\cite{Lopez}.

\begin{figure}[h]
\centering
\includegraphics[scale=0.35]{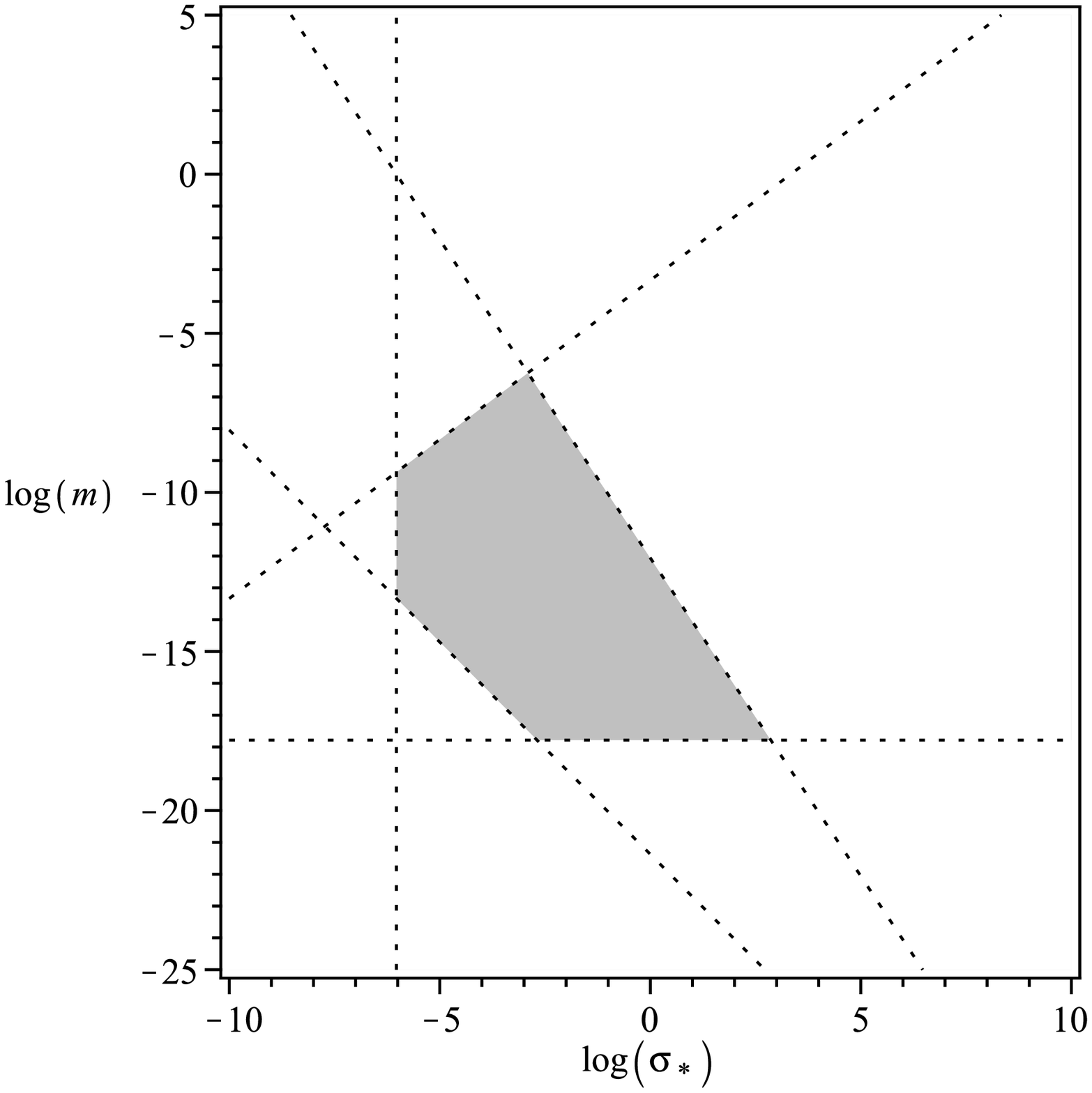}\includegraphics[scale=0.35]{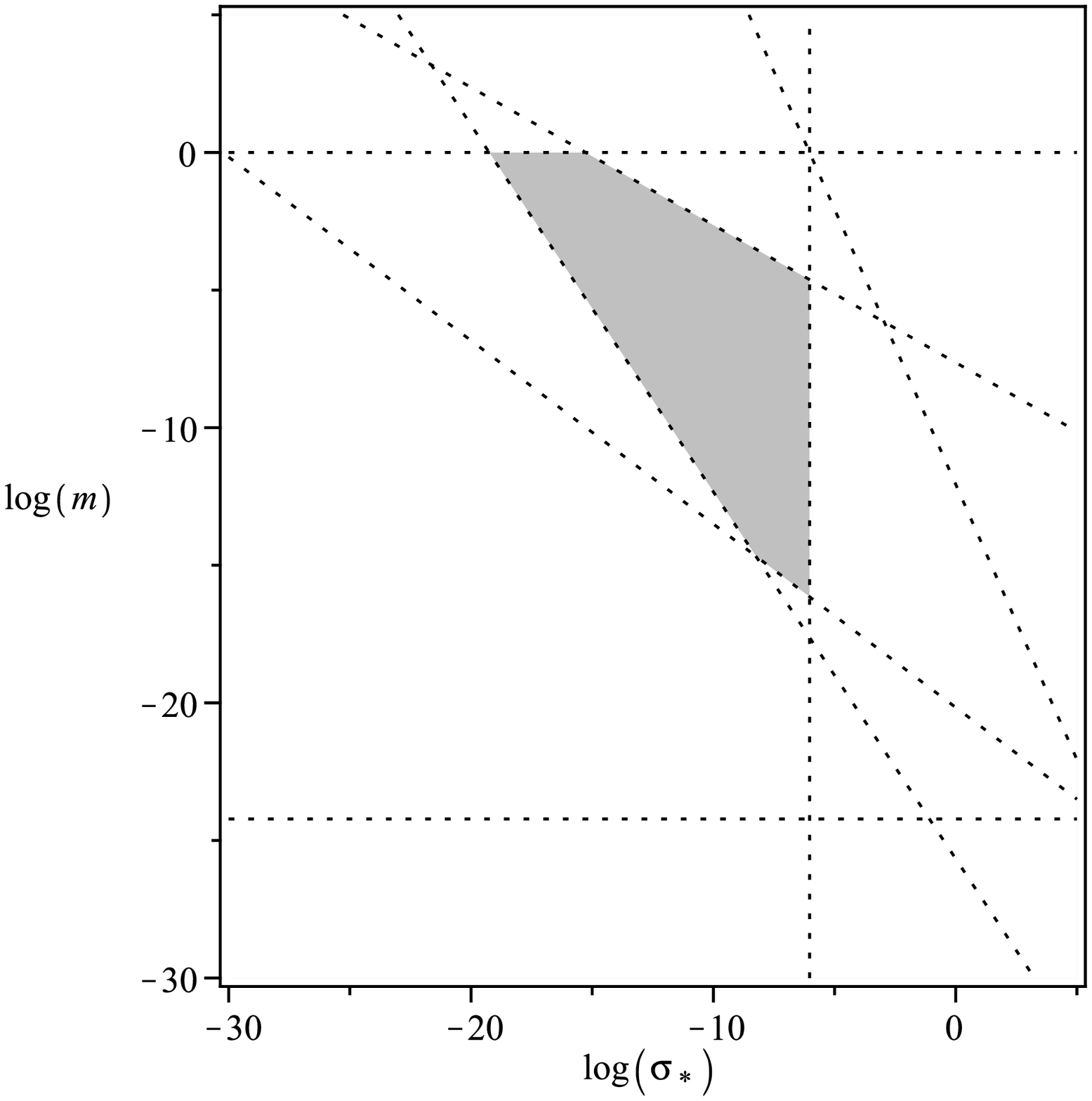}
\caption{Allowed region of parameter space of the curvaton-brane intermediate inflation
model for the case of curvaton domination after decay (left panel) and curvaton domination before decay (right panel). The allowed regions are
shaded.}
\end{figure}

\subsection{Constraint on the reheating temperature}

As in \cite{Nojiri}, the decay rate of curvaton field in some cases can be constrained by,
\begin{equation}
\Gamma_\sigma=g^2m,
\end{equation}
where coupling of curvaton to its decay products is denoted by $g$. Therefore, 
\begin{equation}\label{max}
max(\frac{T_{BBN}}{m^{1/2}},m)\lesssim g \lesssim
min(1,\frac{m\sigma^3}{T_{BBN}^2}),
\end{equation}
is the allowed range for coupling constant. The inequality $m\lesssim g$ is caused by gravitational decay. In the case of curvaton decay before domination, by using $T_{reh}>T_{BBN}$, the
constraint (\ref{max}) represents an upper limit expressed by $g<m\sigma_\ast^3/T_{BBN}^2$.
For the curvaton decay after domination, a lower limit is
given by $T_{BBN}m^{-1/2}<g$.

Employing (\ref{max}) and from \cite{Sanchez}, knowing that $H_\ast\simeq10^{-8}$, $\sigma_\ast\sim10^{-3}$, $m\sim10^{-9}$ (from $n_s\simeq1$), then we get $10^{-9}\lesssim g \lesssim1$. Also, from $T_{reh}\sim gm^{1/2}$, we have $10^{-27/2}\lesssim T_{reh}\lesssim10^{-9/2}$ (in
units of $m_p$)as the allowed range for the reheating
temperature.

Alternatively, by choosing $\sigma_\ast=1$ \cite{Sanchez}, from expression (\ref{max}) the constraint on $g$ becomes $10^{-9}\lesssim g\lesssim1$. Thus, from $T_{reh}\sim gm^{1/2}$, we have the same range for reheating temperature. Note also that the constraint on the density fraction of gravitational waves gives us
$g\sim1$ \cite{Sanchez}. In this case, the reheating temperature
becomes of the order of $T_{reh}\sim10^{-9/2}$ (in units of $m_p$)
which challenges gravitino constraint \cite{Dine}.

\section{Summary}

In this paper, we investigated intermediate
inflation in the brane-world cosmology when curavaton field is liable for universe reheating. We assumed that the universe in both inflationary and kination eras is in high energy regime which is strongly supported by the observation (for example see \cite{del}). In comparison to our previous work, \cite{lowenergy}, where in the kination era the universe was considered to be in low energy regime, new constraints on the space and model parameters are obtained.

Following \cite{del}, two cases have been considered in investigating universe reheating via curvaton mediation in brane cosmology; curvaton dominating the universe after or before its decay. As a result we have obtained an upper limit constrain on $\Gamma_\sigma$ expressed by equation (\ref{gammaafter}) in the case of curvaton dominating the universe after it decays. As is given in equation (\ref{gammabefore}) for the case after decay one obtains a lower limit constraint on $\Gamma_\sigma$. In the intermediate inflationary universe with the scale factor given by equation (\ref{inter}), there are two space parameters $A$ and $f$. In both scenarios, we have acquired constraints for the parameters expresses by equations (\ref{hast}) and (\ref{hast2}). The contour plot of these parameters for different values of e-folding from both lower and upper limits of $T_{BBN}$ is also given in Fig. (1).

We constrained curvaton field and its mass in both cases. One, we obtained six constraints on pair ($m$, $\sigma_\ast$) by the region shown in Fig. 2 (left panel). Similarly, we found seven constraints on them, plotted in Fig. 2 (right panel). Finally, we explore the reheating scenario and from the result of previous sections, we find constraints on the reheating temperature.


\begin{thebibliography}{99}

\bibitem{lowenergy} H. Farajollahi, A. Ravanpak, Can. J. Phys. 88:939-945 (2010)

\bibitem{Tzirakis}K. Tzirakis and W. H. Kinney, JCAP. 01, 028 (2009).

\bibitem{Campo}S. del Campo and R. Herrera, Phys. Lett. B670, 266-270 (2009).

\bibitem{Guth}A. Guth, Phys. Rev. D23, 347 (1981).

\bibitem{Albrecht}A. Albrecht and P. J. Steinhardt, Phys. Rev. Lett. 48, 1220
(1982).

\bibitem{Dunkley}J. Dunkley, E. Komatsu et al., Astrophys. J. Suppl. 180,
306-329 (2009).

\bibitem{Hinshaw}G. Hinshaw, J. L. Weiland et al., Astrophys. J. Suppl. 180, 225-245 (2009).

\bibitem{Rendall}A. D. Rendall, Class. Quant. Grav. 22, 1655-1666 (2005).

\bibitem{Sanyal}A. K. Sanyal, Adv. High Energy Phys. 2009, 630414 (2009).

\bibitem{Barrow}J. D. Barrow, Phys. Lett. B235, 40 (1990).

\bibitem{Muslimov}A. G. Muslimov, Class. Quantum. Grav. 7, 231 (1990).

\bibitem{Liddle}J. D. Barrow and A. R. Liddle, Phys. Rev. D47, 5219-5223 (1993).

\bibitem{Randall}L. Randall and R. Sundrum, Phys. Rev. Lett. 83, 3370-3373 (1999).

\bibitem{Papantonopoulos}E. Papantonopoulos and V. Zamarias, JCAP. 0611, 005 (2006).

\bibitem{Campuzano}C. Campuzano, S. del Campo and R. Herrera, Phys. Rev. D72, 083515 (2005).

\bibitem{Kofman}L. Kofman and A. Linde, JHEP. 0207, 004 (2002).

\bibitem{Felder}G. Felder, L. Kofman and A. Linde, Phys. Rev. D60, 103505
(1999).

\bibitem{del}S. del Campo and R. Herrera, Phys. Rev. D76, 103503 (2007).

\bibitem{Mollerach}S. Mollerach, Phys. Rev. D42, 313 (1990).

\bibitem{Lyth}D. H. Lyth and D. Wands, Phys. Lett. B524, 5 (2002).

\bibitem{Binetruy} P. Binetruy, C. Deffayet and D. Langlois, Nucl. Phys. B. 565, 269 (2000).

\bibitem{Binetruy2} P. Binetruy, C. Deffayet, U. Ellwanger and D. Langlois, Phys. Lett. B. 477, 285 (2000).

\bibitem{Shir} T. Shiromizu, K. Maeda and M. Sasaki, Phys. Rev. D. 62, 024012 (2000).

\bibitem{Lopez}A. R. Liddle and L. A. Urena-Lopez, Phys. Rev. D68, 043517 (2003).

\bibitem{Shiromizu}T. Shiromizu, K. Maeda and M. Sasaki,
Phys. Rev. D62, 024012 (2000).

\bibitem{Kamenshchik}A. Kamenshchik, U. Moschella and V. Pasquier,
Phys. Lett. B511, 265 (2001).

\bibitem{Komatsu}E. Komatsu et al.[WMAP Collaboration], Astrophys. J. Suppl. 180, 330 (2009).

\bibitem{Sanchez}J. C. Bueno Sanchez, K. Dimopoulos, JCAP. 0711, 007 (2007).

\bibitem{Dimopoulos}K. Dimopoulos and D. H. Lyth, Phys. Rev. D69, 123509 (2004).

\bibitem{Langlois}D. Langlois, R. Maartens and D. Wands, Phys. Lett. B489, 259 (2000).

\bibitem{Rodriguez}K. Dimopoulos, D. H. Lyth and Y. Rodriguez, JHEP. 0502, 055 (2005).

\bibitem{Notari}K. Dimopoulos, D. H. Lyth, A. Notari and A. Riotto, JHEP. 0307, 053 (2003).

\bibitem{Dimopoulos2}K. Dimopoulos, Phys. Lett. B634, 331 (2006).

\bibitem{Lazarides}K. Dimopoulos, G. Lazarides, D. Lyth and R. Ruiz de Austri, Phys. Rev. D68, 123515 (2003).

\bibitem{Herrera}C. Campuzano, S. del Campo, R. Herrera, E. Rojas and J. Saavedra,
Phys. Rev. D80, 123531 (2009).

\bibitem{Nojiri}S. Nojiri, S. D. Odintsov and M. Sasaki, Phys. Rev. D71, 123509 (2004).

\bibitem{Dine}M. Dine, Cambridge University Press,
First published in print format (2006).

\end{thebibliography}
\end{document}